\documentclass[%
 aip,
 jmp,%
 amsmath,amssymb,
 reprint,%
]{revtex4-1}

\usepackage{graphicx,natbib}% Include figure files
\usepackage{dcolumn}% Align table columns on decimal point
\usepackage{bm}% bold math

\begin{document}

%\preprint{AIP/123-QED}

\title{Solitonic solutions of Faddeev model}

\author{Chang-Guang Shi}
\email{shicg@shiep.edu.cn}
 \affiliation{College of Mathematics and Physics, Shanghai University of Electric
 Power,Pinglian Road 2103, Shanghai 200090, China}
\author{Minoru Hirayama}%
\affiliation{Department of Physics, University of Toyama, Gofuku
3190, Toyama, Japan}

%\date{}

\begin{abstract}
An application of the equation proposed by the present authors,
which is equivalent to the static field equation of the Faddeev
model, is discussed. Under some assumptions on the space and on the
form of the solution, the field equation is reduced to a non-linear
ODE of second order. By solving this equation numerically, some
solitonic solutions are obtained. It is discussed that the product
of two integers specifying solutions may be identified with the Hopf
topological invariant.
\end{abstract}

\pacs{11.10.Lm,02.30.Ik,03.50-z}

\keywords{Soliton, Nonlinear Field Equation, Faddeev Model}

\maketitle

%%%%%%%%%%%%%%%%%%%%%%%%%%%%%%%%%%%%%%%%%%%%%%%%%%%%%%%%%%%%%%%%%%%%%%%%%

\section {Introduction}

Hopf solitons are classified by the Hopf topological invariant characterizing
the knot structure of system. These topological solutions play
an important role in many areas of physics\cite{J1,J2,J3,J4,J5}. The
minimal model possessing solutions with stable knot structures seems to be the Faddeev
model\cite{Faddeev} which concerns the real scalar fields
\begin{equation}
\boldsymbol{n}(x)=\left(n^1(x),n^2(x),n^3(x)\right)
\end{equation}
 satisfying
\begin{equation}
{\boldsymbol{n}}^2(x)=\boldsymbol{n}(x)\cdot\boldsymbol{n}(x)=\sum\limits_{a=1}^{3}n^a(x) n^a(x)=1
\end{equation}
and is expected to describe the low energy behavior of the $SU(2)$ gauge field\cite{FN}. Although its numerical solutions exhibit quite interesting knot-soliton properties\cite{BS,JP},
the analytic analysis\cite{CM,FL} of the model does not seem to have shown much progress because of the high nonlinearity of the model. \\
The Lagrangian density of the Faddeev model is given by
\begin{align}
{\mathcal L}_F(x)&=c_2 l_2(x)+c_4 l_4(x),\\
l_2(x)&=\partial_{\mu}\boldsymbol{n}(x)\cdot
 \partial^{\mu}\boldsymbol{n}(x),\\
l_4(x)&=-H_{\mu\nu}(x) H^{\mu\nu}(x),\\
H_{\mu\nu}(x)&=\boldsymbol{n}(x)\cdot[\partial_{\mu}\boldsymbol{n}(x)\times
 \partial_{\nu}\boldsymbol{n}(x)]\nonumber\\
&=\epsilon_{abc}n^a(x)\partial_{\mu}n^b(x)\partial_{\nu}n^c(x),
\end{align}
where $c_2$ and $c_4$ are constants.
The field $\bm{n}$ can be expressed by a complex function $u$ as
 \begin{equation}
{\boldsymbol n}=\biggl(\frac{u+u^{*}}{|u|^2+1},\frac{-i(u-u^{*})}{{|u|}^2+1},\frac{{|u|}^2-1}{{|u|}^2+1}\biggr).
\end{equation}
We define $R, \Phi, X$ and $\bm{q}$ by
\begin{align}
R&=|u|,\quad u=R{\rm{e}}^{i\Phi},\\
X&=2\sqrt{\frac{c_4}{c_2}}\frac{1}{1+R^2}=\frac{1}{1+R^2},\\
\bm{q}&=X\nabla u,
\end{align}
where $2\sqrt{c_4/c_2}$ of the dimension of length has been set equal to $1$ and $\bm{q}$ is dimensionless.
If we define a complex $3$-vector $\bm{\alpha}$ and a real $3$-vector $\bm{\beta}$ by\begin{align}
\bm{\alpha}&=\bm{q}^{\star}-\bm{q}^{\star}\times(\bm{q}\times\bm{q}^{\star}),\label{eqn:alphaDEF}\\
\bm{\beta}&=\frac{1}{i}(u^{\star}\bm{q}-u\bm{q}^{\star})= B\nabla
\Phi,~B=\frac{2R^2}{1+R^2},
\end{align}
the static field equation can be written as \cite{HS}
\begin{equation}
\nabla\cdot\bm{\alpha}+i \bm{\beta}\cdot\bm{\alpha}=0
\label{eqn:alphaEq}
\end{equation}
and its complex conjugate. It was found in \cite{HS} that $\bm{\alpha}$ defined by
\begin{align}
\bm{\alpha}&=\nabla \Phi \times \nabla \mu+{\rm{e}}^{-iB\Phi}(\nabla R\times\nabla \nu)\label{eqn:alphaSol}
\end{align}
satisfies (\ref{eqn:alphaEq}) for arbitrary complex functions $\mu$
and $\nu$. Although it is not clear whether the above form of
$\bm{\alpha}$ is general enough or not, we see that the static field
equation of the Faddeev model possesses the above kind of linearity.

%%%%%%%%%%%%%%%%%%%%%%%%%%%%%%%%%%%%%%%%%%%%%%%%%%%%%%%%%

\section{Reduction of Field Equation}

%%%%%%%%%%%%%%%%%%%%%%%%%%%%%%%%%%%%%%%%%%%%%%%%%%%%%
We regard $\mu$ and $\nu$ in (\ref{eqn:alphaSol}) as functions of
$R, \Phi$ and $\zeta$ satisfying $\displaystyle{\frac{\partial(R,
\Phi,\zeta)}{\partial(x_1,x_2,x_3)}\neq 0}$, where $\zeta$ is a real
function. We represent the gradient of $\zeta$ as
\begin{equation}
\nabla \zeta=\Gamma \nabla R\times R\nabla \Phi+\Xi \nabla R+R\Sigma\nabla \Phi
\end{equation}
with $\Gamma, \Xi$ and $\Sigma$ being real functions.
Defining real functions $a,b,c$ and $Y$ by
\begin{align}
\mu_{\zeta}&=\frac{{\rm{e}}^{i(B-1)\Phi}R}{\Gamma}\left[(b+Y)+ia\right],\\
\nu_{\zeta}&=\frac{{\rm{e}}^{i(B-1)\Phi}}{\Gamma}\left[a+i(c+Y)\right],\\
Y&=2X^3
\end{align}
with $\nu_{\zeta}=\frac{\partial\nu}{\partial\zeta}$, etc., we find that the two expressions (\ref{eqn:alphaDEF}) and (\ref{eqn:alphaSol}) for $\bm{\alpha}$ coincide irrespectively of the directions of the $3$-vectors $\nabla R$ and$\nabla\Phi$ if the following relations are satisfied:
\begin{align}
&(\nabla R)^2=\frac{-bX}{a^2-bc},\label{eqn:p}\\
&\nabla R\cdot R\nabla\Phi=\frac{aX}{a^2-bc},\label{eqn:q}\\
&(R\nabla\Phi)^2=\frac{-cX}{a^2-bc},\label{eqn:r}\\
&\nu_{\Phi}+R\Sigma\nu_{\zeta}= {\rm{e}}^{iB\Phi}\left(\mu_{R}+\Xi\mu_{\zeta}\right).\label{eqn:rhomu}
\end{align}
We also obtain the inequalities $b, c\geq 0,~a^2-bc\leq 0$.
On the basis of (\ref{eqn:rhomu}), it was shown\cite{HS} that $a, b$ and $c$ should be determined by the following PDEs:
\begin{align}
\begin{pmatrix}
d_1(\ln \Gamma)\\
d_2(\ln \Gamma)
\end{pmatrix}
=\frac{1}{G}
\begin{pmatrix}
-(c+Y)& a\\
-Ra &R(b+Y)
\end{pmatrix}
\begin{pmatrix}
E\\
F
\end{pmatrix}\label{eqn:lngamma}
\end{align}
with
\begin{align}
&E=(B-1)(c+Y)+D_1(b+Y)-D_2(a),\\
&F=(1-B)a+D_1(a)-D_2(c+Y),\\
&G=R[a^2-(b+Y)(c+Y)],\\
&d_1=\frac{\partial}{\partial R}+\Xi\frac{\partial}{\partial \zeta},~d_2=\frac{\partial}{\partial \Phi}+R\Sigma\frac{\partial}{\partial \zeta},\\
&D_1= R d_1+1+R \Xi_{\zeta},~D_2=d_2+R\Sigma_{\zeta}.
\end{align}
After we find $a,b$ and $c$ satisfying the relation (\ref{eqn:lngamma}), we are left with the first order partial differential equations of the following form:
\begin{align}
\begin{cases}
&(\nabla R)^2=S(R,\Phi,\zeta),\\
&\nabla R\cdot R\nabla \Phi=T(R,\Phi,\zeta),\\
&(R\nabla \Phi)^2= U(R,\Phi,\zeta),
\end{cases}
\end{align}
where $S, T$ and $U$ are the functions fixed by (\ref{eqn:p}),
(\ref{eqn:q}), (\ref{eqn:r}).

In this paper, we consider the  case with a given function
$\zeta(\bm{x})$. The functions $\Gamma, \Xi$ and $\Sigma$ are now
given by
\begin{align}
\Gamma&=\frac{(\nabla R\times\nabla\Phi)\cdot \nabla\zeta}{R(\nabla R\times\nabla\Phi)^2},\label{eqn:Gamma}\\
\Xi&=\frac{[\nabla\Phi\times(\nabla R\times \nabla\Phi)]\cdot \nabla\zeta}{(\nabla R\times\nabla\Phi)^2},\label{eqn:Xi}\\
\Sigma&=\frac{[\nabla R\times(\nabla \Phi\times \nabla R)]\cdot \nabla\zeta}{R(\nabla R\times\nabla\Phi)^2}\label{eqn:Sigma}.
\end{align}

%%%%%%%%%%%%%%%%%%%%%%%%%%%%%%%%%%%
%%%%%%%%%%%%%%%%%%%%%%%%%%%%%%%%%%%%

In the following, we consider the scheme (\ref{eqn:lngamma})
supplemented with (\ref{eqn:Gamma}, \ref{eqn:Xi}, \ref{eqn:Sigma})
to present the example illustrating how the scheme works.

%%%%%%%%%%%%%%%%%%%%%%%%%%%%%%%%%
%%%%%%%%%%%%%%%%%%%%%%%%%%%%%%

\section{Solitonic solutions}

%%%%%%%%%%%%%%%%%%%%%%%%%%%%%%%
%%%%%%%%%%%%%%%%%%%%%%%%%%%%%
It can be seen that all the known exact Hopf solitons of the Nicole model\cite{Nic}
defined by the Lagrangian ${\mathcal L}_N(x)=[ l_2(x)]^{\frac{3}{2}}$
and
the Aratyn-Ferreira Zimerman model\cite{AFZ} defined by
${\mathcal L}_{AFZ}(x)=-[- l_4(x)]^{\frac{3}{4}}$ are solutions of the first order PDEs of the type of (\ref{eqn:alphaSol}) and (\ref{eqn:alphaEq}). \\
We here consider the Faddeev model in the space
\begin{equation}
M=\{\boldsymbol x=(\rho, \varphi, z)~ |~ 2\pi\geq z \geq 0,~2\pi\geq\varphi\geq 0,~\rho_0\geq \rho \geq 0 \},
\end{equation}
where $\rho, \varphi$ and $z$ are cylindrical coordinates.
We impose the boundary condition
\begin{align}
\begin{cases}
&u(\rho, \varphi, 0)=u(\rho, \varphi, 2\pi),\\
&u(\rho,0,z)=u(\rho,2\pi,z),\\
&u(0,\varphi, z)=0,~|u(\rho_0,\varphi, z)|=\infty.
\end{cases}\label{eqn:bc}
\end{align}
We further assume
\begin{align}
\begin{cases}
&R=g(\rho),\\
&\Phi=m\varphi+\ell (z), \quad \ell(0)=0,\quad \ell(2\pi)=2\pi n, \label{eqn:RPhi}
\end{cases}
\end {align}
where $m$ and $n$ are non-vanishing integers.
As for $\zeta$, we adopt a simple choice
\begin{equation}
\zeta=z.
\end{equation}
Anticipating that the nonlinearity of the differential equation for $g(\rho)$
may cause a singularity of $g(\rho)$ at a certain point $\rho_0$,
we have introduced the restriction $\rho_0\geq \rho \geq 0$.
We have also restricted the range of $z$ so that it is an appropriate variable
to descrive the angular part $\Phi$ of $u(\rho,\varphi,z)$. \\
In this case, we have
\begin{equation}
\Gamma=\frac{m\rho}{[m^2+(\ell')^2 \rho^2]gg'},\hspace{1mm} \Xi=0,
\hspace{1mm} \Sigma=\frac{\ell' \rho^2}{[m^2+(\ell')^2 \rho^2]g}
\end{equation}
and
\begin{equation}
a=0,\hspace{1mm} b=\frac{\rho^2}{[m^2+(\ell')^2
\rho^2]g^2(g^2+1)},\hspace{1mm} c=\frac{1}{(g^2+1)(g')^2}
\end{equation}
with $g'=\frac{dg(\rho)}{d\rho}$ and $\ell'=\frac{d\ell(z)}{dz}.$
Then, the relation among $(\frac{\partial}{\partial \rho}, \frac{\partial}{\partial\varphi}, \frac{\partial}{\partial z})$ and
$(\frac{\partial}{\partial R}, \frac{\partial}{\partial\Phi}, \frac{\partial}{\partial\zeta})$ is given by
\begin{align}
&\begin{pmatrix}
\frac{\partial}{\partial\rho}\\
\frac{\partial}{\partial\varphi}\\
\frac{\partial}{\partial z}
\end{pmatrix}
=
\begin{pmatrix}
g'& 0& 0\\
0&m&0\\
0&\ell'&1
\end{pmatrix}
\begin{pmatrix}
\frac{\partial}{\partial R}\\
\frac{\partial}{\partial\Phi}\\
 \frac{\partial}{\partial\zeta}
\end{pmatrix},
\end{align}
or
\begin{align}
\begin{pmatrix}
\frac{\partial}{\partial R}\\
\frac{\partial}{\partial\Phi}\\
\frac{\partial}{\partial\zeta}
\end{pmatrix}
=
\begin{pmatrix}
\frac{1}{g'} & 0 &0\\
0& \frac{1}{m} & 0\\
0& -\frac{\ell'}{m} & 1
\end{pmatrix}
\begin{pmatrix}
\frac{\partial}{\partial \rho}\\
\frac{\partial}{\partial\varphi}\\
 \frac{\partial}{\partial z}
\end{pmatrix},\label{eqn:Trans}
\end{align}
From (\ref{eqn:lngamma}), we have
\begin{align}
&\frac{\partial}{\partial R}{\rm{ln}}\left(\frac{R(b+Y)}{\Gamma}\right)=\frac{1-R^2}{1+R^2}\frac{c+Y}{R(b+Y)},\label{eqn:BY}\\
&\left(\frac{\partial}{\partial\Phi}+R\Sigma \frac{\partial}{\partial\zeta}\right){\rm{ln}\left(\frac{c+Y}{\Gamma}\right)}=-R\Sigma_{\zeta}.\label{eqn:CY}
\end{align}
Noting that $\Gamma$ and $\Sigma$ are independent of $\varphi$, we obtain
\begin{equation}
\frac{\partial}{\partial z}\left(\frac{\Sigma(c+Y)}{\Gamma}\right)=0
\end{equation}
from (\ref{eqn:CY}) and (\ref{eqn:Trans}).
This equality is realized only in the case
\begin{equation}
\ell(z)=nz,\label{eqn:ell}
\end{equation}
from which we obtain $\Phi=m\varphi+nz.$\\
In this case, (\ref{eqn:BY}) becomes the equation to determine $g(\rho)$:
\begin{align}
&\frac{d}{d\rho}\left\{\frac{[2( m^2+n^2 \rho^2) g^2+\rho^2
(g^2+1)^2] g'}{\rho(g^2+1)^3}\right\}\nonumber\\
&=\frac{( m^2+n^2
\rho^2)g(1-g^2)[(g^2+1)^2+2g'^2]}{\rho(g^2+1)^4}.\label{eqn:g}
\end{align}
Through the leading order analysis of (\ref{eqn:g}), we see that the allowed behaviors of $g(\rho)$ near $\rho=0$ and $\rho=\rho_0$ are given by
\begin{align}
&g(\rho)\sim {\rm{const.}}\rho^{|m|}~~ {\rm{or}}~~ g(\rho)\sim {\rm{const.}}\rho^{-|m|} :\rho\sim 0,\\
&g(\rho)\sim \frac{\rm{const.}}{\rho_0-\rho}~~ {\rm{or}}~~g(\rho)\sim {\rm{const.}}(\rho_0-\rho):\rho\lesssim \rho_0.
\end{align}
If there exists a solution satisfying $g(\rho)\sim
{\rm{const.}}\rho^{|m|}: \rho\sim 0$ and $: g(\rho)\sim
\frac{\rm{const.}}{\rho_0-\rho}: \rho\lesssim \rho_0$, it is the
desired one satisfying the boundary condition (\ref{eqn:bc}). The
parameter $\rho_0$ denotes the moving singularity of the solution of
the nonlinear ODE (\ref{eqn:g}). Its value depends on the input,
e.g. $g'(0)$.

In accordance with the behavior $g(\rho)\simeq$ const.$\rho^{|m|}$
near $\rho=0$, we assume $g(0)=0$, $g'(0)=1$ for the case $m=1$ and
$g(0)=g'(0)=0$ for $m=2,3,\cdots$. We show in Fig.1 three examples
of numerical estimation of $g(\rho)$ for the cases $(m, n)=(1,1),
(2,1), (2,2)$. The values of $\rho_0$ in the three cases are given
by 2.34, 0.47 and 0.49, respectively.Conversely, if we fix $\rho_0$,
the value of $g'(0)$ is determined by $m$ and $n$.

%%%%%%%%%%
\begin{figure}
\begin{center}
\includegraphics{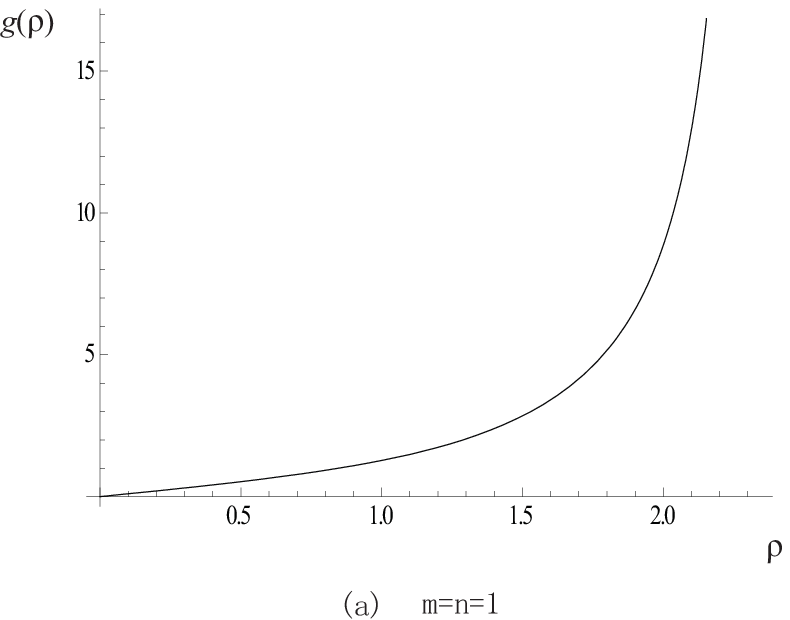}\\
\includegraphics{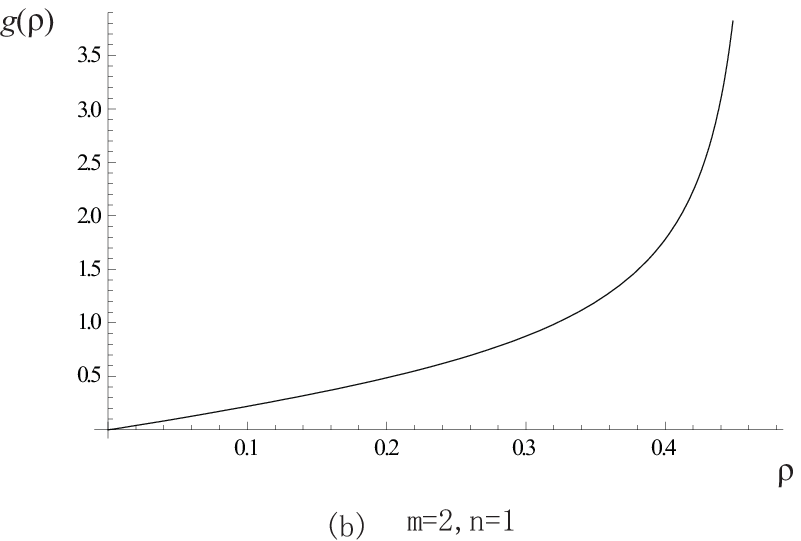}\\
\includegraphics{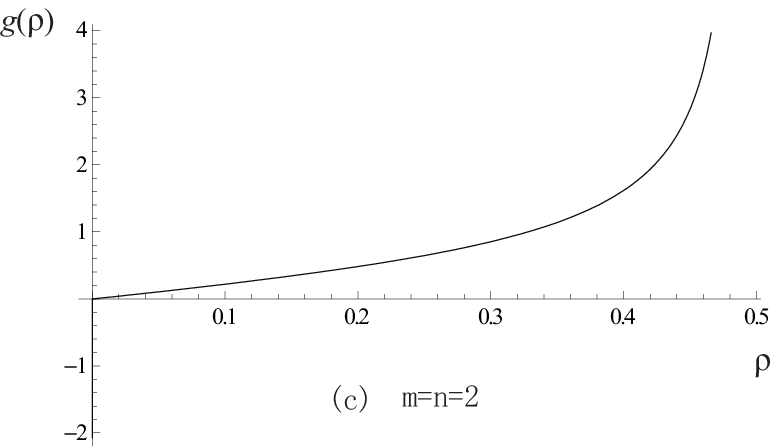}
\end{center}
\caption{Behaviors of $g(\rho)$.}
\end{figure}
%%%%%%%%%%

We note that there exist cases in which we cannot obtain finite and
positive $\rho_0$. In Fig.2, we show the behavior of $g(\rho)$ for
the cases $(m, n)=(1,2),(1,3)$ with $g(0)=0$, $g'(0)=1$. We see that
$g(\rho)$ in these cases is finite for any positive $\rho$.

%%%%%%%%%%
\begin{figure}
\begin{center}
\includegraphics{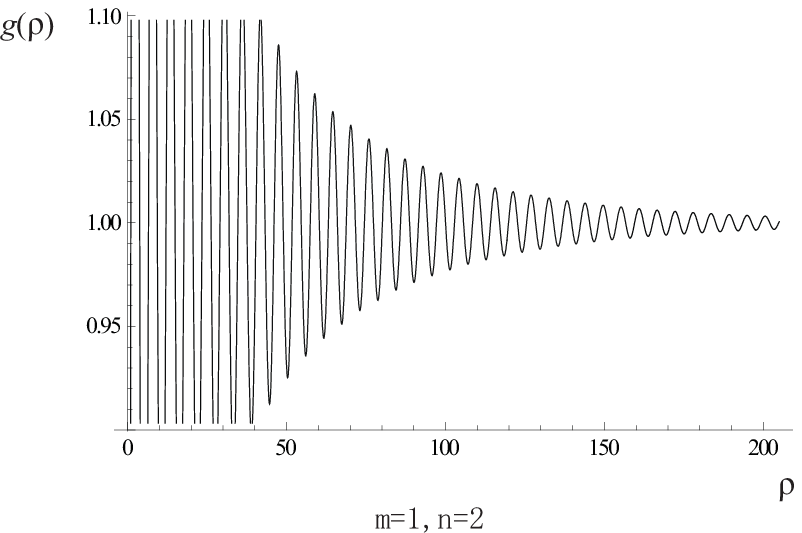}\hspace{3mm}
\includegraphics{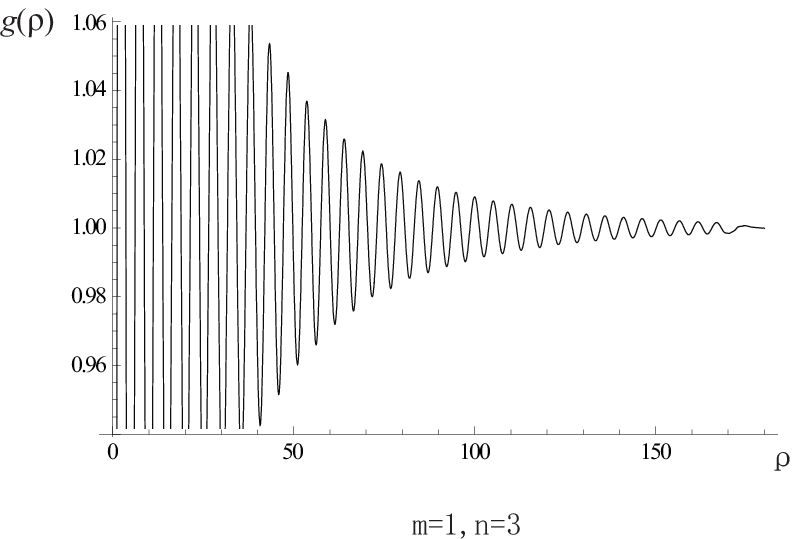}
\end{center}
\caption{Behaviors of $g(\rho)$.}
\end{figure}
%%%%%%%%%%

%%%%%%%%%%
%%%%%%%%%%
\section{Similarity to Hopf soliton}
%%%%%%%%%%%
%%%%%%%%%%%
We now mention on the degree of mapping for $\bm{n}:M \rightarrow S^2$.
We hereafter consider only the case that the conditions $g(0)= 0$ and $g(\rho_{0})=\infty$ are satisfied.
We first define
 $\Phi_{\alpha}(\boldsymbol x)\hspace{1mm} (\alpha=1,2,3,4)$ and $Z(\boldsymbol x)$ by
\begin{align}
&Z(\boldsymbol x)=\binom{Z_1(\boldsymbol x)}{Z_2(\boldsymbol x)},\\
&Z_1(\boldsymbol x)=\Phi_1(\boldsymbol x)+i\Phi_2(\boldsymbol
 x)=\frac{g(\rho)}{\sqrt{1+g(\rho)^2}}{\rm{e}}^{-i n z},\\
&Z_2(\boldsymbol x)=\Phi_3(\boldsymbol x)+i\Phi_4(\boldsymbol
 x)=\frac{1}{\sqrt{1+g(\rho)^2}}e^{im\varphi}.
\end{align}
In terms of $Z(\boldsymbol x)$, the fields $n^a(\boldsymbol x)$ and $u(\boldsymbol x)$ are expressed as
\begin{align}
&n^a(\boldsymbol x)=Z^{\dagger}({\boldsymbol x})\sigma^a Z({\boldsymbol x}),\\
&u(\boldsymbol x)=\biggl({\frac{Z_1(\boldsymbol x)}{Z_2(\boldsymbol x)}}\biggr)^{*},
\end{align}
where $\sigma^a\hspace{1mm}(a=1,2,3)$ are Pauli matrices.
We define the spaces $N_{i} ~ (i=1, 2)$ by $N_{i}= \{ Z_{i}(\boldsymbol x) ~ |~  \boldsymbol x\in M\}$.
The space $M$ is mapped to $N_{1}$ and $N_{2}$ $n$-times and $m$-times, respectively.
Hence $M$ is mapped to the space $N\equiv \{(Z_{1}, Z_{2}) ~ |~  \boldsymbol x\in M\}$
$mn$-times.
This value can be obtained also in the following way.
We define $Q$ classifying the mapping $\bm{n}:M \rightarrow S^2$
 , {\it{\^^ {a} la}} the Hopf charge classifying the mapping $S^3 \rightarrow S^2$, by
\begin{align}
Q&=\frac{1}{16\pi^2}\int_M dV {\boldsymbol
 A}({\boldsymbol x})\cdot{\boldsymbol B}({\boldsymbol x})\nonumber\\
&=\frac{1}{12 \pi^2}\int_M dV \epsilon_{\alpha\beta\gamma\delta}\Phi_{\alpha}\frac{\partial{(\Phi_{\beta},\Phi_{\gamma},\Phi_{\delta})}}{\partial(x,y,z)},\label{eqn:Hopf}\\
&B_i(\boldsymbol x)=\frac{1}{2}\epsilon_{ijk}[\partial_j A_k(\boldsymbol x)-\partial_k A_j(\boldsymbol x)],\\
&A_i({\boldsymbol x})=\frac{1}{i}\{Z^{\dagger}(\boldsymbol x)[\partial_i Z(\boldsymbol x)]-[\partial_i Z^{\dagger}(\boldsymbol x)]Z(\boldsymbol x)\},
\end{align}
where $\epsilon_{\alpha\beta\gamma\delta}$ is the four-dimensional
Levi-Civita symbol satisfying $\epsilon_{1234}=1$.
From (\ref{eqn:Hopf}), it is straightforward to obtain
\begin{eqnarray}
Q= -mn(\frac{1}{[g(\rho_0)]^2+1}-\frac{1}{[g(0)]^2+1})=mn.
\end{eqnarray}
In the conventional discussion of Hopf solitons,
$\bm{n}$ satisfying $\bm{n(\bm{x})}=$ const. for $~|~ \bm{x} ~|~=\infty$
is regarded as a field defined in the three-dimensional Euclidean
space with the points at $~|~ \bm{x} ~|~= \infty$ identified,
which is isomorphic to $S^{3}$.
In our case, recalling the conditions
$u(\rho, \varphi, 0)= u(\rho, \varphi, 2\pi)$,
$u(\rho,0,z)= u(\rho,2\pi,z)$ and $g(\rho_{0})= \infty$,
our $\bm{n}$ may be regarded as a field on a solid torus
with the points on the surface $\rho=\rho_{0}$ identified,
which we denote by $\mathcal{M}$.
Thus, we have seen that the mapping $\bm{n}: \mathcal{M}\rightarrow S^{2}$
can be classified similarly to the mapping $\bm{n}: S^{3}\rightarrow S^{2}$.

%%%%%%%%%%%%%%%%%%%%%%%%%%%%%%%%%%%%%%%%%

\section{Summary }
%%%%%%%%%%%%%%%%%%%%%%%%%%%%%%%%%%%%%%%%%%%%
We considered an application of the equation
for the 3-component vector
$\bm{\alpha}=\bm{q}^{\star}-\bm{q}^{\star}\times(\bm{q}\times\bm{q}^{\star})$,
which is equivalent to the static field equation of the Faddeev model.
Under some special assumptions, the equation was reduced to a nonlinear ODE of second order.
Solutions of this equation was investigated numerically.
The similarity to the case of Hopf soliton was discussed.

\begin{acknowledgments}
This research was partially supported by the Innovation Program of
Shanghai Municipal Education Commission£¨Grant No. 09ZZ183£©and the
Natural Science Foundation of Shanghai(Grant No. 11ZR1414100).
\end{acknowledgments}


\begin{thebibliography}{99}
\bibitem{J1}J.E. Moore, 2010 {\it Nature} {\bf 464} 194.
\bibitem{J2}Y. Xia et al, 2009 {\it Nature Physics} {\bf 5} 398.
\bibitem{J3}D. Hsieh, et al., 2009 {\it Science} {\bf 323} 919.
\bibitem{J4}Y. L. Chen et al. 2009 {\it Science} {\bf 325} 178.
\bibitem{J5}J. E. Moore, Y. Ran,
X.-G. Wen, 2008 {\it Phys. Rev. Lett.} {\bf 101} 186805.
\bibitem{Faddeev}  L. Faddeev, 1976
{\it Lett. Math. Phys. } {\bf 1} 289.
\bibitem{FN} L. Faddeev  and A. J. Niemi,  1999
{\it Phys. Rev. Lett. } {\bf 82} 1624.
\bibitem{BS} R. A. Battye and P. M. Sutcliffe,   1998
{\it Phys. Rev. Lett. } {\bf 81} 4798.
\bibitem{JP}J. Hietarinta  and  P. Salo, 1999 {\it Phys. Lett. B} {\bf~451} 60.
\bibitem{CM} C-G Shi  and M. Hirayama,  2008 {\it Int.J.Mod.Phys.A} {\bf 23} 1361.
\bibitem{FL} L. A. Ferreira,  2009 {\it JHEP} 0905, 001.
\bibitem{HS}M. Hirayama  and C-G Shi, 2007 {\it Phys.Lett.B} {\bf~652} 384.
\bibitem{Nic} D. A. Nicole, 1978 {\it J. Phys. C} {\bf~4} 1363.
\bibitem{AFZ}H. Aratyn , L. A. Ferreira  and A. H. Zimerman,  1999 {\it Phys. Lett. B} {\bf~456} 162.
\end{thebibliography}
\end{document}